\documentclass[prb,twocolumn,showpacs,superscriptaddress]{revtex4}
\usepackage[dvips]{epsfig}
\usepackage{graphicx}
\usepackage{amsfonts}
\usepackage{bm}
\usepackage{amsmath,amssymb,lscape,float}
 
\def\tr{\qopname\relax{no}{Tr}}

\newcommand{\ket}[1]{{\vert #1 \rangle}}
\newcommand{\bra}[1]{{\langle #1 \vert}}
\newcommand{\dg}{\dagger}

\newcommand{\ga}{{\alpha}}
\newcommand{\gb}{{\beta}}

\newcommand{\gpsi}{{\psi}}
\newcommand{\gPhi}{{\Phi}}

\makeatletter
\renewcommand\appendix{\par
  \setcounter{section}{0}%
  \setcounter{subsection}{0}%
  \setcounter{equation}{0}%
  \gdef\thesection{\Alph{section}}
  \@addtoreset {equation}{section}
  \renewcommand{\theequation}{\thesection\arabic{equation}}}
\makeatother

\begin{document}

\title{Quantum-entanglement aspects of polaron systems}
\author{Vladimir M. Stojanovi\'c\footnote[2]{Electronic mail: vstojano@andrew.cmu.edu}}
\affiliation{Department of Physics, Carnegie Mellon University,
             Pittsburgh, Pennsylvania $15213$, USA}

\author{Mihajlo Vanevi\'c}
\affiliation{Departement Physik, Universit\"{a}t Basel,
 Klingelbergstrasse 82, CH-4056 Basel, Switzerland}
\affiliation{School of Physics, Georgia Institute of Technology,
Atlanta, Georgia 30332, USA}

\date{\today}

\begin{abstract}
We describe quantum entanglement inherent to the polaron ground states of
coupled electron-phonon (or, more generally, particle-phonon) systems
based on a model comprising both local (Holstein-type)
and nonlocal (Peierls-type) coupling. We study this model using a variational method
supplemented by the exact numerical diagonalization on a system of finite size.
By way of subsequent numerical diagonalization of the reduced density matrix, we
determine the particle-phonon entanglement as given by the von Neumann
and linear entropies. Our results are strongly indicative of the intimate relationship
between the particle localization/delocalization and the particle-phonon entanglement. In particular,
we find a compelling evidence for the existence of a nonanalyticity in the entanglement entropies
with respect to the Peierls-coupling strength. The occurrence of such nonanalyticity --
not accompanied by an actual quantum phase transition -- reinforces analogous conclusion drawn
in several recent studies of entanglement in the realm of quantum-dissipative systems. In addition,
we demonstrate that the entanglement entropies saturate inside
the self-trapped region where the small-polaron states are nearly maximally mixed.
\end{abstract}
\maketitle

\section{Introduction}
In recent years entangled quantum systems~\cite{PlenioReview:07,EntangleBook}
have garnered interest as a resource for quantum information processing.~\cite{Bennett:95}
In addition, a great deal of research effort has been expended towards
clarifying the role of entanglement in the (zero-temperature) quantum
phase transitions of many-particle systems.~\cite{Osterloh+Nielsen:02,Galindo:02,LidarSham:06}
Perhaps the most compelling, however, is the need to elucidate its possible bearing on
the macroscopic properties of physical systems.~\cite{vedralovi,Factorization}
Namely, while the entanglement entropies are simply related to the many-body density
matrix of the system, they bear no {\em a priori} relation to any observable
physical quantity. Attempts to associate a physical meaning with entanglement
were made, for example, in mesoscopic physics: schemes have been proposed
for detection~\cite{Choi:00} and even measurement of
entanglement by exploiting its emergent relation
to the quantum noise.~\cite{mesoentangle,BeenakkerLec:06}

Central to all the above developments is the problem of quantifying
entanglement in diverse physical systems, such as quantum spin
chains,~\cite{Verstraete:04,Juancho:04,Fan:04} interacting bosons
and/or fermions,~\cite{Brandao:05,Larsson:05,Vollbrecht:07,Banuls:08}
quantum-dissipative~\cite{stauber+guinea,LeHur:08} and disordered systems,~\cite{Kopp:07}
to name but a few.~\cite{Amico:08} In this regard, one of the areas of
condensed-matter physics whose quantum-entanglement aspects have heretofore
received only scanty consideration is that of the polaron problem. Ever since its
inception by Landau and Pekar,~\cite{landaupekar} the polaron
concept~\cite{firsovbook,AlexandrovBook,PolaronReview}-- a quantum
particle interacting with a bosonic environment -- has played an
immensely important role in theoretical studies of coupled electron--
or exciton--phonon (henceforth e-ph) systems.~\cite{polaroncrucial}
What is more, this truly ubiquitous concept is lately finding
resurgence in seemingly unrelated physical situations,
the realm of ultracold atoms being a case in point.~\cite{ColdPolaron}

The main body of polaron-related work is focussed on the study
of a single electron interacting with the harmonic
lattice vibrations through a short-range, non-polar potential that
is linear in the lattice displacements and describes the
dependence of the electronic on-site energies on the lattice
degrees of freedom. The traditional starting point in describing
such interaction, dubbed local e-ph coupling, is the
paradigmatic molecular-crystal model due to Holstein.~\cite{polaroncrucial}
As is being amply appreciated lately,~\cite{polaronhtc,Perroni:04,Perroni:05}
however, local coupling is not the only type of short-range e-ph interaction
relevant in realistic systems: nonlocal (off-diagonal) e-ph coupling accounts
for the phonon-modulation of the electronic hopping integrals and bears relevance
to several classes of molecule-based systems. The most common form of nonlocal coupling
is Peierls-type coupling,~\cite{Zaanen:94,zaanendictaat,yonemitsu:07} widely studied
within the framework of the semi-classical Su-Schrieffer-Heeger (SSH) model~\cite{SSH,ZoliIn,zolicouple}
that describes the anomalous transport properties of nonlinear excitations (solitons, polarons)
along the quasi-one-dimensional polyacetilene chain. In an implicit way -- becoming manifest
by carrying out the Jordan-Wigner transformation -- this coupling forms the basis of
the $XY$ spin-Peierls model,~\cite{Cross:79,Barford:05} hence the name.
Likewise, coupling to the breathing mode in cuprate superconductors is of Peierls' type and
-- as transpires from recent investigations based on generalizations of the $t-J$ model --
plays an important role in these systems, especially in the regime of weak doping.~\cite{polaronhtc}
Somewhat different forms of nonlocal coupling have been
shown to be of relevance for charge transport in organic molecular
crystals,~\cite{hannewald:04,stojanovic:04} carbon nanotubes,~\cite{Torres:06}
and DNA wires.~\cite{dnaschoen:07} It is worth of mention that a form of nonlocal coupling
has also been incorporated in a generic electron-boson coupling model recently
proposed by Alvermann {\em et al.}.~\cite{alvermann:07}

Given the abiding interest in the polaron problem in condensed-matter
physics, the quantum-entanglement aspects of this problem have so far not
been given due attention. While the changeover from a small to a
large polaron is known to have the nature of a smooth crossover (with no broken symmetry),
rather than a phase transition,~\cite{Gerlach:87} it is still tempting to quantify it
using the entanglement measures. Entanglement in the one-dimensional Holstein
model was studied by Zhao {\em et al.}~\cite{zhaozanardi} The authors emphasized the
relation between the self-trapping process and the quantum (hetero-) entanglement
between the phonon subsystem and electronic excitation. Effects of Peierls-type
coupling, however, are as yet totally unexplored; given the wealth of
intriguing implications of such interaction it appears
interesting to elucidate the role it plays in particle localization as seen through
entanglement measures. Additional motivation comes
from recent investigations of entanglement in the realm of quantum dissipative
systems.~\cite{stauber+guinea,LeHur:08}
One of the most important conclusions of these studies is that entanglement measures
can have non-analyticities away from any phase transition and that these
non-analyticities are intimately related to the loss of coherence.
To be more specific, Stauber and Guinea~\cite{stauber+guinea} found a
nonanalyticity at the transition from underdamped to overdamped oscillations
in the Ohmic case~\cite{Porras:08} of the spin-boson model. Moreover, this
nonanalyticity proved to be even more pronounced than the one occurring at the
actual localization phase transition.

As a matter of fact, an evidence that non-analytic behavior of entanglement-related
quantities does not necessarily coincide with the quantum phase transitions had already been
found before.~\cite{Osterloh+Nielsen:02}
Besides, a study of localizable entanglement in a gapped quantum spin
system~\cite{Verstraete:04} has showed that entanglement length diverges despite the
fact that the correlation length remains finite, the latter indicating absence of a
quantum phase transition. On the other hand, in disordered systems, for example,
no such cases have been reported. It is therefore of interest to
investigate whether the polaron problem -- somewhat related to the spin-boson model,
but with no phase transitions taking place -- may also defy the tenet whereby
an occurrence of a nonanalyticity in the entanglement entropy is a telltale
signature of a quantum phase transition.

In the present work, we study entanglement in polaron
systems. In order to address the problem from as general a viewpoint
as possible, we start from a polaron model that includes both local
(Holstein-type) and nonlocal (Peierls-type) short-range e-ph coupling.
Given that this Hamiltonian does not admit an exact solution in any physically-relevant limit,
we analyze it using a variational method supplemented by an exact diagonalization on a
finite-size system. Proceeding in this way, we find that entanglement exhibits non-analyticities
as a function of Peierls-type coupling, which are not accompanied by a phase transition.
We argue that the occurrence of such ``accidental'' nonanalyticities in the problem
at hand is related to the loss of coherence, much like in the quantum-dissipative systems.

The outline of the remainder of this paper is as follows.
In Sec.~\ref{model} we present the model and notation
to be used throughout. Sec.~\ref{variat} contains details of our
variational approach: we first introduce our variational Ansatz,
then provide details of analytical derivations needed to implement it,
and finally describe our computational method of variational
minimization. The following Sec.~\ref{exactdiag} contains essential
details of the exact diagonalization method. In Sec.~\ref{entang} we introduce the
entanglement measures and lay out the method for calculating them.
The obtained results are presented in Sec.~\ref{res}, accompanied by
a discussion of their salient features. We conclude, with some
general remarks, in Sec.~\ref{concl}.

\section{Model} \label{model}

The system under study consists of an excess particle (electron, hole,
exciton) interacting with harmonic lattice vibrations
-- dispersionless (Einstein) phonons -- through a short-range
interaction. As our starting point, we adopt a one-dimensional
e-ph Hamiltonian obtained by dovetailing
the Peierls-type coupling term on the conventional Holstein
Hamiltonian. (Restriction to a one-dimensional system is not a severe
limitation given the short-range nature of the e-ph interactions discussed 
here.) The compact form of this extended Holstein model reads
\begin{multline}\label{}
\hat{H}=\sum_{i}\varepsilon_{i}(\{\hat{\mathbf{u}}\})\:
\hat{a}_{i}^{\dagger}\hat{a}_{i}+
\sum_{i} t_{i+1,i}(\{\hat{\mathbf{u}}\})\:
(\hat{a}_{i+1}^{\dagger}\hat{a}_{i}+
\mathrm{h.c.})\\
+\omega\sum_{i}\:\hat{b}_{i}^{\dagger} \hat{b}_{i} \:,
\end{multline}
\noindent where $\hat{a}_{i}^{\dagger}$($\hat{a}_{i}$) creates
(destroys) a particle at $i$-th site (at position
$R_{i}$,\:$i=0,1,\ldots, N-1$), $\hat{b}_{i}^{\dagger}$($\hat{b}_{i}$)
creates (destroys) a dispersionless phonon with frequency $\omega$ at
the same site. The effective on-site energy
\begin{equation}\label{}
\varepsilon_{i}(\{\hat{\mathbf{u}}\})=\varepsilon
+\ga_{\scriptscriptstyle H}\hat{u}_{i} \:,
\end{equation}
\noindent and hopping integral
\begin{equation}\label{}
t_{i+1,i}(\{\hat{\mathbf{u}}\})= -t+
\ga_{\scriptscriptstyle P}(\hat{u}_{i+1}-\hat{u}_{i}) \:,
\end{equation}
\noindent depend on the lattice displacements
$\hat{u}_{i}\equiv(2M\omega)^{-\scriptscriptstyle 1/2}
(\hat{b}_{i}+\hat{b}_{i}^{\dagger})$,  where $\ga_{\scriptscriptstyle H}$ and
$\ga_{\scriptscriptstyle P}$ are the local (Holstein-type) and nonlocal
(Peierls-type) coupling constants, respectively, and $M$ is the mass of
molecules in the underlying crystal. The bare on-site energy and hopping integral
are denoted by $\varepsilon$ and $t$, respectively. For simplicity we
take $\varepsilon=0$ in what follows. More explicitly, the Hamiltonian
can be written as
\begin{equation} \label{Hamiltonian}
\hat{H} = \hat{H}_{\rm e} + \hat{H}_{\rm ph} + \hat{H}_g + \hat{H}_{\phi} \:,
\end{equation}
\noindent where
\begin{align}
\hat{H}_{\rm e} &= -t \sum_i (\hat{a}^\dg_{i+1}\hat{a}_i + \mathrm{h.c.}) \:,
\\
\hat{H}_{\rm ph} &= \omega \sum_i \hat{b}^\dg_i \hat{b}_i \:,
\\
\hat{H}_{g} &= g \omega \sum_i \hat{a}^\dg_i \hat{a}_i (\hat{b}^\dg_i + \hat{b}_i) \:,
\\
\hat{H}_{\phi} &= \phi \omega \sum_i (\hat{a}^\dg_{i+1}\hat{a}_i + \mathrm{h.c.})
(\hat{b}^\dg_{i+1} + \hat{b}_{i+1} - \hat{b}^\dg_i - \hat{b}_i) \:,
\end{align}
\noindent with $g\equiv\ga_{\scriptscriptstyle H}/\sqrt{2M\omega^3}$
and $\phi\equiv\ga_{\scriptscriptstyle P}/\sqrt{2M\omega^3}$
being the dimensionless local and nonlocal coupling constants,
respectively. The two important limiting cases of our model
are the Holstein model ($\phi=0$) and the quantized version of the
SSH model ($g=0$).

The eigenstates of Hamiltonian \eqref{Hamiltonian} ought to be
the good-quasi-momentum states, i.e., eigenstates of the total crystal
momentum operator
\begin{equation}\label{totalcryst}
\hat{K}=\sum_{k} k\:\hat{a}^{\dagger}_{k}
\hat{a}_{k}+\sum_{q}q\:\hat{b}^{\dagger}_{q}
\hat{b}_{q} \:\:,
\end{equation}
\noindent since the latter commutes with $\hat{H}$.
In the following, the eigenvalues of $\hat{K}$ will be labelled with $\kappa$. By
making use of the Born-von Karman periodic boundary conditions, the quasi-momenta in the
first Brillouin zone are given by $\kappa=(2\pi/a) (m/N)$
($a$ -- the lattice spacing;\:$m=0,1,\ldots, N-1$).
For convenience, we express quasi-momenta in units of $a^{-1}$,
so that $\kappa R_n=\kappa na \to \kappa n$.

\section{Variational method} \label{variat}
\subsection{Choice of variational Ansatz}

While not being exactly-soluble, Hamiltonian in Eq.~\eqref{Hamiltonian}
can be treated variationally; in the studies of coupled e-ph systems
methods of this type have been shown to yield quantitatively trustworthy results
that compare well with those obtained by the exact (numerical)
diagonalizations.~\cite{Weisse:00,Ku+Bonca:02}
An important class of such methods is furnished by Toyozawa's Ansatz
state~\cite{Toyozawa:61} and generalizations
thereof.~\cite{Romero+Lindenberg,Ku+Bonca:02,Barisic:02} While these
Ansatz states have been widely used in the studies of the Holstein
model, they are also capable of describing
systems with simultaneous local and nonlocal coupling.

Omitting the most conventional form of Toyozawa's Ansatz state, we purposefully
cast it in a way that renders manifest its entangled nature:
\begin{equation}\label{ansatz}
|\psi_{\kappa}\rangle=\frac{1}{N}\sum_{n,m}
e^{i\kappa n}\gamma^{\kappa}_{m-n}\:\hat{a}_{m}^{\dagger}
|0\rangle_{\textrm{e}}\otimes \prod_{l}
\hat{D}_{l}(\xi_{l-n}^{\kappa})|0\rangle_{\textrm{ph}} \:,
\end{equation}
\noindent where $|0\rangle_{\textrm{e}}$ ($|0\rangle_{\textrm{ph}}$) is
the electron (phonon) vacuum, and $\hat{D}_{l}(\alpha)\equiv
\exp(\alpha\hat{b}_{l}^{\dagger}-\alpha^{*}\hat{b}_{l})$
is Glauber's displacement operator that creates the phonon
coherent state $|\xi_{l-n}^{\kappa}\rangle_{\textrm{ph}}\equiv
\hat{D}_{l}(\xi_{l-n}^{\kappa})|0\rangle_{\textrm{ph}}$ at site $l$.
This (overcomplete) set of phonon coherent states captures the
multi-phononic nature of the polaron ground state. Importantly, it
is the dependence of the variational parameters $\gamma_{m-n}^{\kappa}$ on both
$m$ and $n$ that renders Ansatz state $|\psi_{\kappa}\rangle$ entangled:
the sums over these indices cannot be decoupled, implying that this state
cannot be expressed as a separable (direct-product) state in the Hilbert
space ${\mathcal H}={\mathcal H}_{\textrm{e}}\otimes{\mathcal
H}_{\textrm{ph}}$.

Generally speaking, the use of variational methods invariably involves the
trade-off between flexibility of the variational wave-function (which
increases with the growing number of parameters) and numerical difficulty
of finding reliably the global minimum of the ground-state-energy expectation
value (complexity grows rapidly with parameter number; see Sec.~\ref{compproc}
for additional details). In this regard, the major drawback of Toyozawa's Ansatz is
that it involves a large number -- $2N$ for each $\kappa$ -- of variational
parameters. Alternative methods have been proposed that provide accurate results,
while involving smaller number of parameters. Adopting this point of view, we
seek the polaron eigenstates of Hamiltonian \eqref{Hamiltonian} in the form
of translationally-invariant Bloch states
\begin{equation}\label{trialfunc}
\ket{\gpsi_\kappa} = \frac{1}{\sqrt{N}}
\sum_n e^{i\kappa n}\:\ket{\gpsi_\kappa(n)} \:,
\end{equation}
\noindent with ``form-factors'' $\ket{\gpsi_\kappa(n)}$
given by~\cite{Perroni:04}
\begin{multline}\label{formfact}
\ket{\gpsi_\kappa(n)}=\sum_m \gPhi_\kappa(m)\;
e^{i\kappa m}\:\hat{a}^\dg_{n+m} \ket{0}_{\rm e}
\\
\otimes \exp\Big(\sum_{j=-2}^{2}\hat{U}_{\kappa;j}(n+j)
\Big)\ket{0}_{\rm ph}\:.
\end{multline}
\noindent The skew-Hermitian operators $\hat{U}_{\kappa;j}$ ($j=0,\pm 1,\pm 2$)
are defined as
\begin{equation}
\hat{U}_{\kappa;j}(n) = \frac{1}{\sqrt{N}}
\sum_q \Big(f_{\kappa;j}(q)\;e^{iqn}\;
\hat{b}_q -\textrm{h.c.}\Big)\:,
\end{equation}
\noindent with
\begin{equation}
f_{\kappa;j}(q)=\frac{\ga_{\kappa;j}}%
{1+2\;(t/\omega)\;\gb_{\kappa;j}[\cos(\kappa)-\cos(\kappa+q)]} \:.
\end{equation}

Using trial wave-functions \eqref{trialfunc}, the lowest polaron
band $E_{\textrm{\tiny{GS}}}(\kappa)$ can be obtained by minimizing the
energy expectation value over variational parameters
$V_{\kappa}=\big\{\ga_{\kappa;\scriptscriptstyle 0},
\ga_{\kappa;\scriptscriptstyle\pm 1},
\ga_{\kappa;\scriptscriptstyle \pm 2};
\gb_{\kappa;\scriptscriptstyle 0},
\gb_{\kappa;\scriptscriptstyle \pm 1},
\gb_{\kappa;\scriptscriptstyle \pm 2};
\gPhi_{\kappa}(-5),\ldots,\gPhi_{\kappa}(5) \big\}$:
\begin{equation}\label{gsexpect}
E_{\textrm{\tiny{GS}}}(\kappa)=\min_{\displaystyle V_{\kappa}}
\:\frac{\langle\psi_{\kappa}|\:\hat{H}\:|\psi_{\kappa}\rangle}
{\langle\psi_{\kappa}|\psi_{\kappa}\rangle} \:.
\end{equation}
\noindent [Note that for each $\kappa$ there are in total twenty variational
parameters; this number is fixed, rather than being proportional to
the system size as in Toyozawa's Ansatz. Importantly, among eleven parameters
$\gPhi_{\kappa}(j)$ only ten are independent because of the normalization
condition on the trial wave-function in Eq. \eqref{trialfunc}.] In the following,
we will be particularly interested in the polaron ground-state energy
$E_{\textrm{\tiny{GS}}}(\kappa=0)\equiv E_{0}$. The corresponding
variationally-optimized ground-state wave-function will hereafter be denoted
with $|\textrm{GS}\rangle$.

\subsection{Matrix elements and computational scheme} \label{compproc}

Here we present the derivation of the expression for
the energy expectation value, followed by the details of
our numerical method for variational minimization.
To facilitate further derivations we first note that
\begin{equation}\label{}
\exp\Big(\sum_{j=-2}^{2}\hat{U}_{\kappa;j}(n+j)\Big)=
\prod_{q}\hat{D}_{q}\Big(-\frac{e^{-iqn}w_{\kappa}^{*}(q)}
{\sqrt{N}}\Big) \:,
\end{equation}
\noindent where $\hat{D}_{q}(\alpha_{q})\equiv \exp(\alpha_{q}
\hat{b}_{q}^{\dagger}-\alpha_{q}^{*}\hat{b}_{q})$ is Glauber's displacement
operator that creates a coherent state of phonons with quasi-momentum $q$,
and $w_{\kappa}(q)$ is defined as
\begin{equation}
w_{\kappa}(q)\equiv\sum_{j=-2}^{2}f_{\kappa;j}(q)\:e^{iqj} \:.
\end{equation}
\noindent It is straightforward to show that
\begin{equation}\label{psipsi}
\langle \gpsi_\kappa \vert \gpsi_\kappa \rangle =
\sum_{mm'}\gPhi_{\kappa}^{*}(m')\gPhi_{\kappa}(m)\;
Z^{\kappa}_{m-m'} \:,
\end{equation}
\noindent where
\begin{equation} \label{zdef}
\begin{array}{rl}
&\displaystyle Z^{\kappa}_{m-m'}\equiv \\
&\displaystyle_{\textrm{\tiny{ph}}}\Big\langle 0\Big|
\prod_{q}\hat{D}_{q}^{\dagger}\Big(-\frac{e^{-iqm}w_{\kappa}^{*}(q)}
{\sqrt{N}}\Big)\hat{D}_{q}\Big(-\frac{e^{-iqm'}w_{\kappa}^{*}(q)}
{\sqrt{N}}\Big)\Big|0\Big\rangle_{\textrm{\tiny{ph}}} \:.
\end{array}
\end{equation}
\noindent By making use of the well-known expression~\cite{Scully+Zubairy}
for the overlap of coherent states $\langle\alpha|\beta\rangle$
($\alpha,\beta \in \mathbb{C}$)
\begin{equation}\label{}
\bra{\alpha} \beta \rangle=\langle 0|\:\hat{D}^{\dagger}
(\alpha)\hat{D}(\beta)\:|0\rangle=e^{\alpha^{*}\beta}\:
e^{-\frac{1}{2}(|\alpha|^{2}+|\beta|^{2})}
\end{equation}
\noindent we obtain
\begin{equation}
Z^{\kappa}_{m-m'}=\exp\Big[-\frac{1}{N} \sum_q
\Big(1-e^{iq(m-m')}\Big)|w_{\kappa}(q)|^2\Big] \:.
\end{equation}
\noindent Other relevant matrix elements are given by
\begin{multline}
\bra{\gpsi_\kappa}\; \hat{H}_{\rm e}\; \ket{\gpsi_\kappa}
= -t\sum_{mm'}\gPhi_{\kappa}^{*}(m')\gPhi_{\kappa}(m)
\\
\times\Big(e^{-i\kappa}\:Z^{\kappa}_{m-m'+1}+e^{i\kappa}
\:Z^{\kappa}_{m-m'-1}\Big) \:,
\end{multline}
%
\begin{multline}\label{hphon}
\bra{\gpsi_\kappa}\; \hat{H}_{\rm ph}\; \ket{\gpsi_\kappa}
= \frac{\omega}{N}\sum_{mm'}\gPhi_{\kappa}^{*}(m')
\gPhi_{\kappa}(m)\;Z^{\kappa}_{m-m'}
\\ \times \sum_q e^{iq(m-m')}\; |w_{\kappa}(q)|^{2} \:,
\end{multline}
%
\begin{multline}\label{hg}
\bra{\gpsi_\kappa}\; \hat{H}_{g} \; \ket{\gpsi_\kappa}
=-\frac{g\omega}{N}\sum_{mm'}\gPhi_{\kappa}^{*}(m')
\gPhi_{\kappa}(m)\;Z^{\kappa}_{m-m'}
\\
\times \sum_q \Big(e^{-iqm'}w_{\kappa}(q)+e^{iqm}w_{\kappa}^{*}(q)\Big) \:,
\end{multline}
%
\begin{multline}\label{hphi}
\bra{\gpsi_\kappa}\; \hat{H}_{\phi} \; \ket{\gpsi_\kappa}
=\frac{\phi\omega}{N} \sum_{mm'}
\gPhi_{\kappa}^{*}(m')\gPhi_{\kappa}(m)\sum_q \\
\times\Big(e^{i\kappa}\; Z^{\kappa}_{m-m'-1}(1-e^{-iq})
-e^{-i\kappa}\; Z^{\kappa}_{m-m'+1}(1-e^{iq})\Big)\\
\times\Big(e^{-iqm'}w_{\kappa}(q)-e^{iqm}w_{\kappa}^{*}(q)\Big) \:.
\end{multline}
\noindent The above formulae are easily derived using identity
\begin{equation}\label{}
e^{\hat{A}}e^{\hat{B}}=e^{\hat{A}+\hat{B}}
e^{\frac{1}{2}[\hat{A},\hat{B}]} \:,
\end{equation}
\noindent which holds if operators $\hat{A}$ and $\hat{B}$ satisfy
condition $[\hat{A},[\hat{A},\hat{B}]]=[\hat{B},[\hat{A},\hat{B}]]=0$,
as well as identity
\begin{equation}\label{}
\big[\:\hat{b},f(\hat{b}^{\dagger},\hat{b})\:\big]=
\frac{\partial f(\hat{b}^{\dagger},\hat{b})}
{\partial\hat{b}^{\dagger}} \:,
\end{equation}
\noindent valid for an arbitrary analytic function
$f(\hat{b}^{\dagger},\hat{b})$ of bosonic operators.

An important measure of the multi-phononic nature of the
polaron ground state is the average number of phonons
\begin{equation}\label{avephonon}
\bar{N}_{\textrm{ph}}=\:\Big\langle
\textrm{GS}\Big|\:\sum_{i}\hat{b}_{i}^{\dagger}\hat{b}_{i}\Big|
\textrm{GS}\Big\rangle \:,
\end{equation}
\noindent which can be obtained using Eq. \eqref{hphon}.

Based on the expressions for $\langle\psi_{\kappa}|\:\hat{H}\:|\psi_{\kappa}\rangle$
and $\langle\psi_{\kappa}|\psi_{\kappa}\rangle$, we perform variational minimization
in order to find the polaron ground state ($\kappa=0$) for a system with $N=32$ sites.
Due to the large number ($n=20$) of variational parameters involved, the
energy expectation value is a function of these parameters with multiple local minima.
Finding the global minimum thus constitutes a rather nontrivial numerical optimization
problem. We perform this complex task using the multi-start-based global random search
method:~\cite{OptimBook} we first generate a large sample ($\sim 10^{5}$) of random
points in the space of variational parameters; we then select a smaller
number ($\sim 20$) of them that have the smallest values of the function
to be minimized and perform local searches for minima
[$O(n^{3})$ computation] around each of these points: the one with the smallest
energy is then adopted as the sought-after global minimum. The fidelity of this approach
is corroborated by the stability of the final result for the global minimum upon
varying the initial number of random points.

Regarding the choice of our numerical method a remark is in order here. In the
local-coupling-only case ($\phi=0$, i.e., the Holstein model),
the system is exactly soluble in the ``unphysical'' limit of zero hopping
($t\rightarrow 0$) by the well-known Lang-Firsov canonical transformation.~\cite{polaroncrucial}
The natural way to proceed in finding the optimal variational parameters for finite $t$
is then to start from the exact solution for $t=0$ and gradually change $t$, using optimal
values of variational parameters obtained for given value of $t$ as the initial guess for
the next, slightly higher value. However, for finite $\phi$ the model is not exactly-soluble
in any relevant limit. Therefore, the procedure just described does not carry over to the $\phi\neq 0$
case and one needs a careful sampling of the entire space of variational parameters, afforded by the
multi-start-type methods, to reliably find the global minimum of the ground-state-energy
expectation value.

\subsection{Application scope of our variational method and comparison
            with other methods}

\begin{figure}[h!]
\hspace*{-7.5mm}%
\includegraphics[scale=1.375]{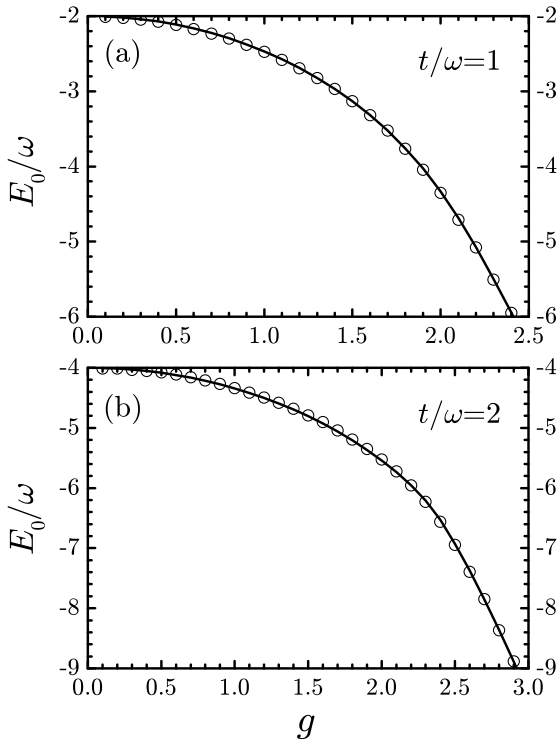}
\caption{\label{compare}
Comparison of the ground-state energies obtained using the variational
method of Cataudella {\em et al.}~\cite{Cataudella:99} (circles)
and the method of the present work (solid curve) in the $\phi=0$
case: $t/\omega=1.0$ (a) and $t/\omega=2.0$ (b).}
\end{figure}

While Ansatz state in Eqs.~\eqref{trialfunc}-\eqref{formfact}
was originally introduced for use in the Peierls-coupling-only case (quantized SSH model),
we here demonstrate the that it can also be utilized for the local-coupling-only case
(Holstein model), and accordingly for the case with simultaneous local and nonlocal
couplings. The nearly-perfect agreement of our results in the $\phi=0$ case with another
known variational approach, suggested by Cataudella {\em et al.}~\cite{Cataudella:99},
is illustrated in Fig.~\ref{compare}a,b. The latter was shown to agree well with the
Global-Local method~\cite{Romero+Lindenberg} implying the general agreement of all three methods.
Therefore the trial states used here are generally applicable to short-ranged
e-ph interactions with Einstein phonons. The agreement between the results
obtained by the variational methods of the present type and other approaches
(density-matrix renormalization group method,~\cite{Jeckelmann:98} quantum Monte
Carlo~\cite{PolaronQMC}) is also well established.~\cite{Romero+Lindenberg}

\section{Exact diagonalization} \label{exactdiag}

To supplement our variational method, we also perform
an exact numerical diagonalization of Hamiltonian \eqref{Hamiltonian}
on a system of finite size. We study system of $N=6$ sites, varying
the maximal number of phonons between $M=8$ and $M=10$. The states
in the truncated Hilbert space are given by
\begin{equation}
|\Psi\rangle=\sum_{\mathbf{n},\mathbf{m}}C_{\mathbf{n,m}}
\:|\mathbf{n}\rangle\otimes|\mathbf{m}\rangle \:,
\end{equation}
\noindent where $\mathbf{n}=(n_{\scriptscriptstyle 0},...,
n_{\scriptscriptstyle N-1})$ and $\mathbf{m}=(m_{\scriptscriptstyle 0},...,
m_{\scriptscriptstyle N-1})$ are the sitewise electron and
phonon occupation numbers ($\sum_{i}n_{i}=1,\:\sum_{i}m_{i}\leq M$).
Coefficients $C_{\mathbf{n,m}}$ contain the information about the phonon
content of state $|\Psi\rangle$.
\begin{figure}[h!]
\hspace*{-2.5mm}%
\includegraphics[scale=0.75]{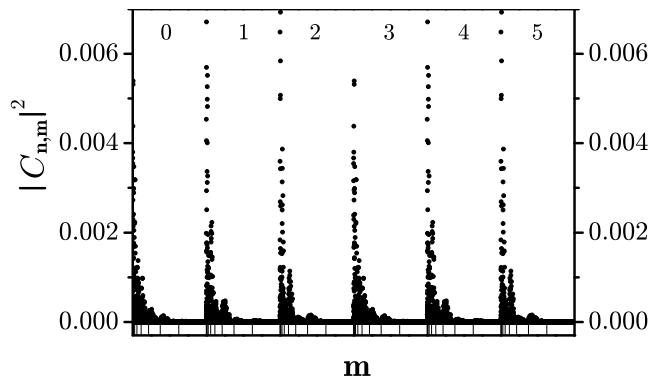}
\caption{\label{phonondistr}
Phonon distribution sitewise in the polaron ground state
for $t/\omega=1.0$, $\phi=1.3$ ($N=6$, $M=10$). The states
$\mathbf{m}$ with the same total number of
phonons $m=\sum_{i}m_{i}$ are indicated by the bins
at the bottom portion of the plot.}
\end{figure}

One of the crucial prerequisites for a successful application of the exact-diagonalization
approach in the present context is a proper truncation of the (otherwise infinite-dimensional)
phonon Hilbert space. In other words, the maximal total number of phonons on a lattice
has to be large enough as to be capable to account for the phonon distribution
in the polaron ground state in the strong-coupling regime.
Typical phonon content (sitewise) of the polaron ground state
obtained by exact diagonalization is illustrated in Fig.~\ref{phonondistr},
where the $n$-th group of peaks represents the phonon distribution for the case
when the electron is located at site $n$ ($n=0,...,N-1$). Within each group of peaks,
coefficients $C_{\mathbf{n},\mathbf{m}}$ are given in the ascending order of the phonon
occupation numbers. For weak coupling, phonon distribution
$p_m=\sum_{\mathbf{m}}|C_{\mathbf{n},\mathbf{m}}|^2$
(where $m=\sum_i m_i$) peaks at $m=0$ phonons, while in the strong-coupling regime
it peaks at $m=3$ or $4$ phonons. This {\em a posteriori}
corroborates that our choice of the maximal number of phonons
(between $8$ and $10$) was pertinent.

\section{Entanglement} \label{entang}
\subsection{Entanglement entropies: generalities}

In order to make the present work self-contained, before embarking on the
calculation of entanglement in our coupled e-ph system we review the general
prescription for characterization of bipartite quantum entanglement.
We consider a composite quantum system that can be divided up into two
parts A and B, where A denotes the subsystem of interest and B the
environment whose details are unimportant. The Hilbert space of the full system
has the form of a tensor product: $\mathcal{H}=\mathcal{H}_{\textrm{\tiny{A}}}
\otimes\mathcal{H}_{\textrm{\tiny{B}}}$. In a pure state $|\Psi\rangle$,
the density matrix of the full system is given by
\begin{equation}\label{rhogen}
\hat{\rho}=\frac{|\Psi\rangle\langle\Psi|}{\langle\Psi|
\Psi\rangle} \:.
\end{equation}
\noindent We then construct the reduced (marginal) density matrix
$\hat{\rho}_{\textrm{\tiny{A}}}$ by tracing over the environmental
degrees of freedom: $\hat{\rho}_{\textrm{\tiny{A}}}
=\tr_{\textrm{\tiny{B}}}\hat{\rho}$.
The von Neumann (entanglement) entropy, defined by
\begin{equation}
S = -\textrm{Tr}_{\textrm{\tiny{A}}}(\hat{\rho}_{\textrm{\tiny{A}}}
\ln\hat{\rho}_{\textrm{\tiny{A}}}) \:,
\end{equation}
\noindent contains information about the quantum correlations
present in the pure quantum state under study. It represents the most widely used
measure of bipartite quantum entanglement. Note that $S=
-\tr_{\textrm{\tiny{A}}}(\hat{\rho}_{\textrm{\tiny{A}}}\ln\hat{\rho}_{\textrm{\tiny{A}}})=
-\tr_{\textrm{\tiny{B}}}(\hat{\rho}_{\textrm{\tiny{B}}}\ln\hat{\rho}_{\textrm{\tiny{B}}})$,
where the reduced density matrix $\hat{\rho}_{\textrm{\tiny{B}}}$ is
obtained by tracing over the degrees of freedom in subsystem A. The upper bound
$S_{\textrm{max}}=\ln(D)$ of $S$, where $D$ is dimensionality of the reduced density matrix,
is reached when the reduced density matrix is maximally mixed.

The so-called linear entropy is defined as
\begin{equation}
S_{\scriptscriptstyle L}=
1-\textrm{Tr}_{\textrm{\tiny{A}}}
(\hat{\rho}^{2}_{\textrm{\tiny{A}}}) \:,
\end{equation}
\noindent and has the advantage (compared to the von Neumann entropy)
of being easier to calculate. It is worth
of mention that both the von Neumann and linear entropies are closely
related to the quantum R\'{e}nyi entropies~\cite{EntangleBook}
$S_{q}(\hat{\rho})\equiv(1-q)^{-1}\ln(\textrm{Tr}\hat{\rho}^{q})$ ($q\geq 0$):
$S(\hat{\rho})$ is the $q\rightarrow 1$ limit of $S_{q}(\hat{\rho})$, while
$S_{\scriptscriptstyle L}(\hat{\rho})$ is simply related to $S_{2}(\hat{\rho})$.
The upper bound of $S_{\scriptscriptstyle L}$, reached for the maximally-mixed reduced
density matrix, is given by $S_{{\scriptscriptstyle L},\textrm{max}}=1-D^{-1}$.

\subsection{Reduced density matrix from variational approach}

In accordance with general relation \eqref{rhogen},
the density matrix corresponding to the state $|\psi_{\kappa}\rangle$
on the tensor-product Hilbert space ${\mathcal H}={\mathcal H}_{\textrm{e}}
\otimes{\mathcal H}_{\textrm{ph}}$ is given by
\begin{equation}\label{}
\hat{\rho}_{\textrm{e-ph}}(\kappa)=\frac{|\psi_{\kappa}\rangle
\langle\psi_{\kappa}|}{\langle\psi_{\kappa}|\psi_{\kappa}\rangle} \:.
\end{equation}
\noindent The reduced particle (electron) density
matrix is given by the partial trace over the phonon Hilbert space
${\mathcal H}_{\textrm{ph}}$:
\begin{equation}
\hat{\rho}_{\textrm{e}}(\kappa)= \textrm{Tr}_{\textrm{ph}}
\big[\hat{\rho}_{\textrm{e-ph}}(\kappa)\big] \:.
\end{equation}
\noindent Straightforward derivation, with the aid of
Eq.~\eqref{zdef}, yields
\begin{equation}\label{trsumfinal}
\begin{array}{rl}
&\displaystyle \textrm{Tr}_{\textrm{ph}}
\big(\ket{\gpsi_{\kappa}}\bra{\gpsi_{\kappa}}\big)=\frac{1}{N}
\sum_{nn'}\:\Big(\sum_{m,m'}\Phi_{\kappa}^{*}(m')\Phi_{\kappa}(m) \\
&\displaystyle \times \:Z^{\kappa}_{m-m'+n'-n}\Big)\: e^{i\kappa(n-n')}
\ket{n}_{\textrm{e}}\:_{\textrm{e}}\bra{n'} \:,
\end{array}
\end{equation}
\noindent where $\ket{n}_{\textrm{e}}\equiv
a^{\dagger}_{n}|0\rangle_{\textrm{e}}$ is the state with electron
at site $n$. The last equation, when combined with Eq.~\eqref{psipsi},
readily leads to the expression for the general matrix element of
the reduced density matrix $\hat{\rho}_{\textrm{e}}(\kappa=0)\equiv
\hat{\rho}_{\textrm{e}}$ in the polaron ground state:
\begin{equation}\label{reddensmatzero}
\displaystyle
\big(\hat{\rho}_{\textrm{e}}\big)_{nn'}=\frac{1}{N}
\frac{\sum_{\scriptscriptstyle mm'}\Phi_{\kappa\scriptscriptstyle=0}^{*}(m')
\Phi_{\kappa\scriptscriptstyle=0}(m)\;Z^{\kappa\scriptscriptstyle=0}_{m-m'+n'-n}}
{\sum_{\scriptscriptstyle mm'}\gPhi_{\kappa\scriptscriptstyle=0}^{*}(m')
\gPhi_{\kappa\scriptscriptstyle=0}(m)\;Z^{\kappa\scriptscriptstyle=0}_{m-m'}} \:.
\end{equation}

The corresponding von Neumann entropy
\begin{equation}\label{vonneuman}
S=-\textrm{Tr}_{\textrm{e}}
(\hat{\rho}_{\textrm{e}}\ln\hat{\rho}_{\textrm{e}})
\end{equation}
\noindent can be expressed as
\begin{equation}\label{vonentrop}
S=-\sum_{i}\lambda_{i}\ln\lambda_{i} \:,
\end{equation}
\noindent where $\left\{\lambda_{i}\:|\: i=0,1,\ldots, N-1\right\}$
are the eigenvalues ($\lambda_{i}>0\:,\:\sum_{i}\lambda_{i}=1$) of
$\hat{\rho}_{\textrm{e}}$. Based on our variational approach, the
linear entropy
\begin{equation}\label{linentropy}
S_{\scriptscriptstyle L}=
1-\textrm{Tr}_{\textrm{e}}(\hat{\rho}_{\textrm{e}}^{2})
\end{equation}
\noindent can readily be obtained in an analytical form using
Eq. \eqref{reddensmatzero}; however, we here omit the ensuing
cumbersome expression. In the exact-diagonalization
approach, the reduced density matrix $\hat{\rho}_{\textrm{e}}$ is
obtained from the corresponding eigenvectors.
The results illustrating dependence of the entanglement
entropies on the e-ph coupling strengths are presented
in the following section.

\section{Results and Discussion} \label{res}

\begin{figure}[h!]
\hspace*{-7.5mm}%
\includegraphics[scale=1.675]{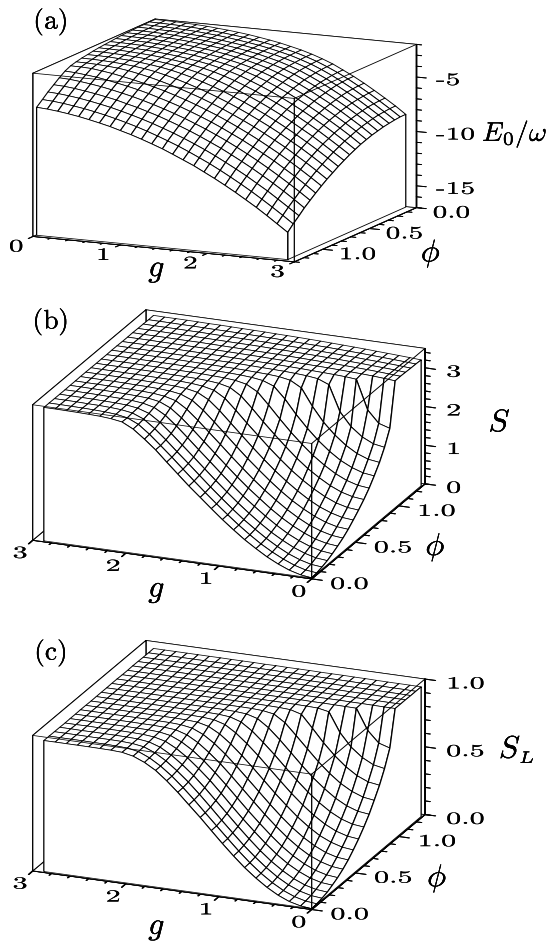}
\caption{\label{fig:ES_vs_g_phi}
Variational ground-state energy (a), the von Neumann entropy (b), and linear
entropy (c) as functions of dimensionless local ($g$) and nonlocal
($\phi$) coupling strengths, for $t/\omega=1.0$ and $N=32$.}
\end{figure}

Our numerical results, obtained using variational approach, correctly reproduce a continuous
dependence of the polaron ground-state energy, displayed in Fig.~\ref{fig:ES_vs_g_phi}a for
$t/\omega=1.0$, on local and nonlocal coupling strengths, reflecting the well-known
absence of phase transitions in coupled e-ph systems.~\cite{Gerlach:87}

The entanglement entropies, depicted in Fig.~\ref{fig:ES_vs_g_phi}b,c
for $t/\omega=1.0$, both behave in a similar way: they increase with
increasing coupling strengths, saturating and remaining essentially
unchanged in the self-trapped region where the particle becomes
localized. The saturation values of the two entropies
are close to those of the maximally-mixed density matrix:
$S_{\textrm{max}}=\ln(N)$, $S_{{\scriptscriptstyle L},\textrm{max}}=1-N^{-1}$.
For example, for $N=32$, $S_{\textrm{max}}=3.46$,
$S_{{\scriptscriptstyle L},\textrm{max}}=0.97$
(cf. Fig. \ref{fig:ES_vs_g_phi}).

\begin{figure}[t!]
\hspace*{-7.5mm}%
\includegraphics[scale=1.7]{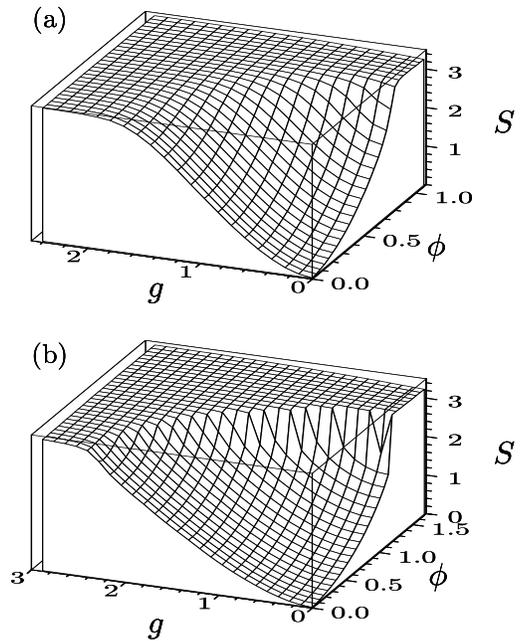}
\caption{\label{twoentropies}
The von Neumann entropy $S$, obtained variationally, as a function
of dimensionless local ($g$) and nonlocal ($\phi$) coupling strengths. Displayed
are results for $t/\omega = 0.25$ (a) and $t/\omega=2.0$ (b), with $N=32$.}
\end{figure}

Importantly, for $t/\omega\gtrsim 0.85$ we find a strong nonanalyticity
in the dependence of entanglement entropies on the nonlocal coupling strength.
This nonanalyticity has the character of a jump-discontinuity and is more pronounced 
in the case of von Neumann entropy (Fig.~\ref{fig:ES_vs_g_phi}b) than for the linear
entropy (Fig.~\ref{fig:ES_vs_g_phi}c). It becomes more and more pronounced
with increasing value of $t/\omega$, i.e., upon approaching the adiabatic
regime $t\gg \omega$. In order to emphasize that discontinuous behavior sets in
for sufficiently large value of the ratio $t/\omega$, in Fig.~\ref{twoentropies}a
we depict the von Neumann entropy for $t/\omega=0.25$ where the nonanalyticity
does not exist at all, and for $t/\omega=2.0$ (Fig.~\ref{twoentropies}b) where the
nonanalyticity is noticeably more pronounced than for $t/\omega=1.0$.

\begin{figure}[t!]
\hspace*{-3.5mm}%
\includegraphics[scale=2.2]{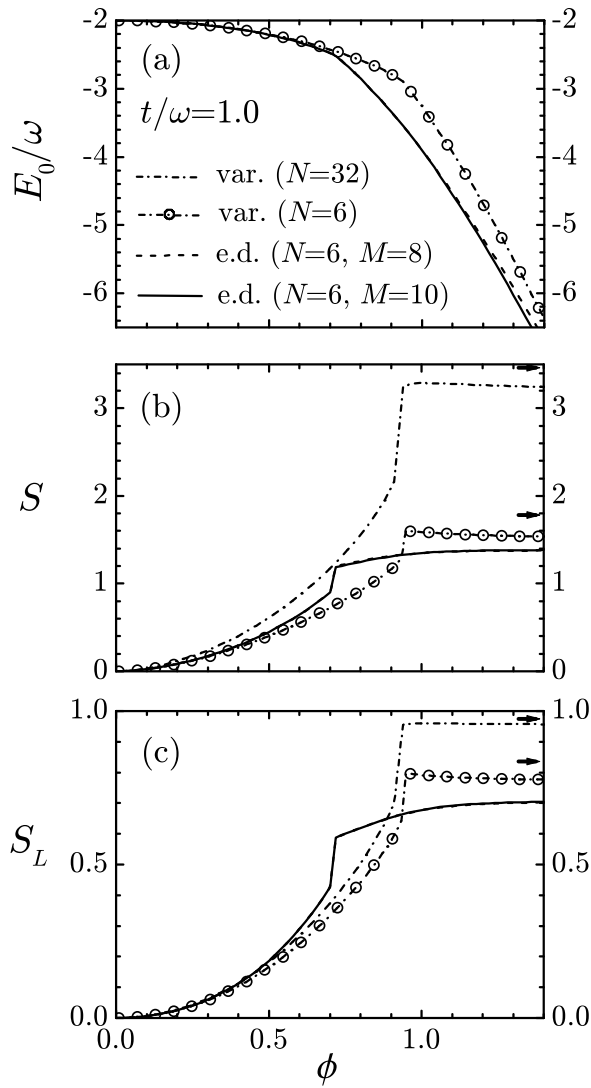}
\caption{\label{2Dphi}
Ground-state energy (a), the von Neumann entropy (b), and linear entropy
(c) for $t/\omega=1.0$ and $g=0$, as calculated using the variational (var.)
and exact diagonalization (e.d.) approaches. $N$ is the number of sites,
while $M$ stands for the maximal number of phonons used for exact
diagonalization. The arrows indicate the entropies for the maximally-mixed
reduced density matrix.}
\end{figure}

To emphasize the appearance of a nonanalyticity in the entanglement
entropies as a function of nonlocal coupling strength, we study the nonlocal-coupling-only
case making comparisons between the results of the two approaches (variational
and exact diagonalization). Typical results are depicted in Fig.~\ref{2Dphi}.
Figure \ref{2Dphi}a illustrates very good agreement between the two approaches
as far as the ground-state energy is concerned. In Figs.~\ref{2Dphi}b,c the entanglement
entropies are shown as obtained from two different variational calculations
(for $N=6$ and $N=32$) and two different exact-diagonalizations (with $N=6$ and
maximum $M=8$ or $10$ phonons used). The only sizeable discrepancy is
in the behavior of $S$, which clearly stems from the finite-size effects:
while values of $S$ corresponding to the variational calculation with $N=32$ deviate
considerably from those of exact diagonalizations with $N=6$, the difference between the
two approaches when applied to a system with the same number of sites is not very drastic.
However, the most important feature of the obtained results, manifest in all the cases
considered, is the non-analytic behavior of $S$ and $S_{\scriptscriptstyle L}$ with respect
to $\phi$.

Detailed analysis shows that the observed nonanalyticities occur for
values of $\phi$ at which the lowest energy states of $\hat{H}(\phi)$ undergo
avoided crossings. The crossings are avoided (rather than real ones)
because $[\hat{H},\hat{H}_{\phi}]\neq 0$,~\cite{sachdevbook} leading
to a smooth dependence of the polaron ground-state energy
on the coupling strength $\phi$ (and, accordingly, the absence of a phase
transition in the conventional sense of the term). There is, however, no general
principle that would rule out the occurrence of nonanalyticities in the
entanglement entropies at these avoided-crossing points.

The fact that nonanalyticities of the type discussed here do not exist in the
case with local e-ph coupling -- nor in the ``statically''-disordered systems described by the
Anderson model -- and that for Peierls-type coupling
they show up only when $t$ is larger or of the same order as $\omega$
point to the possible importance of ``dynamical disorder'' (retardation, i.e., nonlocality in time)
effects, which are here bringing about the nonlocal particle-phonon correlations.
Namely, when the relevant electron energy scale (set by the hopping integral $t$)
becomes comparable to or larger than the characteristic phonon
energy ($\omega$), the lattice deformation does not follow instantaneously
the electron motion. Consequently, the phonon
modes that are excited by the passage of the electron take a
long time to relax. Therefore, a lattice deformation can be
observed far away from the current position of the electron.
As a consequence, the effects of retardation in the e-ph interaction become
prominent. Such effects are known to be much more pronounced for the Peierls-type
interaction than for the purely local Holstein-type interaction, even when
the effects of phonon dispersion are accounted for in the latter.~\cite{zolicouple} This
can be traced back to the fact that unlike local coupling, which is momentum-independent,
the Peierls-type coupling depends strongly on both the electron and phonon
momenta. More precisely, in momentum-space this coupling reads
\begin{equation}\label{mscoupling}
\hat{H}_{\phi}=\frac{1}{\sqrt{N}}\sum_{k,q}\gamma(k,q) \:
\hat{a}_{k+q}^{\dagger}\hat{a}_{k}
(\hat{b}_{-q}^{\dagger}+\hat{b}_{q}) \:,
\end{equation}
\noindent where $\gamma(k,q)$ is the e-ph interaction vertex function
\begin{equation}\label{vertex}
\gamma(k,q)=i\:\frac{2\alpha_{\scriptscriptstyle P}}{\sqrt{2M\omega}}
\:\Big(\sin(k)-\sin(k+q)\Big) \:.
\end{equation}
\noindent In particular, at small phonon momenta $\gamma$
behaves as
\begin{equation}\label{lowq}
\gamma(k,q)\propto \:q  \qquad  (\:q\rightarrow 0\:) \:,
\end{equation}
\noindent which is a very strong momentum dependence, and
different than that of the Fr\"{o}hlich e-ph interaction.~\cite{PolaronReview}
Localization of a ``Peierls polaron'' is therefore expected to have a much more
dramatic impact on the nature of the accompanying e-ph correlations than that
of a ``Holstein polaron''. Additionally,
the more pronounced character of the nonanalyticity observed for larger $t/\omega$
also appears to be in consistency with this argument; it demonstrates the
increasing ``inertia'' of the more and more spatially-extended phonon cloud to
electron's localization.

In the light of our findings and those of Stauber and Guinea,~\cite{stauber+guinea} it is tempting
to draw some parallels between the two models involved, or more specifically,
between our model and 'Ohmic systems'. However, our polaron model -- at least in its full
form -- does not seem to bear a direct relation to any of the known quantum-dissipative
models, because of the intrinsically off-diagonal nature in the electron bilinear operators
and the dispersionless character of phonons that it involves. [Besides, based on Eq.~\eqref{lowq}
we can infer that the Peierls'-coupling term with acoustic (rather than optical) phonons
would be the most similar to the super-Ohmic systems with spectral density
$J(\omega)\propto\omega^{2}$,
even though we are here concerned with a one-dimensional system.]
The link to these models appears much easier to establish for the local-coupling-only
Holstein model ($\phi=0$): the two-site version of this model, discussed long time ago
by Shore and Sander,~\cite{Shore:73} represents a simplistic (single-mode) 
form of the spin-boson model.
However, the nonanalyticities of entanglement entropies
that we find in the present work occur in the polaron
crossover regime, and are accompanied by the growth in the
average number of phonons in the polaron ground state 
[cf. Eq.~\eqref{avephonon}].
This can indeed be considered as a physical situation analogous
to the loss of coherence in the spin-boson-type models.

A few remarks regarding our variational method are in order.
While examples are known of artifacts~\cite{Chen+Wong:08pre}
in the variational approaches to coupled e-ph or quantum-dissipative systems -- a
prominent one being the failure of variational methods to predict a continuous
transition in the sub-Ohmic case of the spin-boson model -- in the case at hand such an approach
correctly reproduces the smooth dependence of the polaron ground-state energy on both local
and nonlocal coupling strengths (the polaron crossover). The obtained nonanalyticities in
the entanglement entropies, corroborated through the exact diagonalizations,
represent a robust feature that underscores the connection between entanglement
and localization.

\section{Conclusions} \label{concl}

In summary, we have investigated the quantum-entanglement aspects
of polaron systems. As our point of departure, we have adopted
a very general polaron model that includes both local (Holstein-type)
and nonlocal (Peierls-type) particle-phonon coupling. We have studied
this Hamiltonian using a sophisticated variational approach supplemented by
the exact (numerical) diagonalization on a finite-size system. We have established a 
close connection between the entanglement and phonon-induced localization. Our
results make it transparent that the entanglement entropies constitute much more 
sensitive indicators of the change of polaron states than the ground-state
energy. While the intuition as to the relationship between entanglement and 
localization has already been manifest from studies of other classes of 
systems~\cite{Brand:08} -- most prominently the
disordered ones -- our findings add to it some other elements.

As a salient feature, we have demonstrated that -- above some threshold value
for the ratio of the hopping integral and the phonon frequency -- entanglement entropies
exhibit a nonanalyticity as a function of the nonlocal (Peierls)
coupling strength. This nonanalyticity is physically related to the loss of coherence
and is not accompanied by a phase transition. In this sense, the present work
reinforces the conclusions drawn in some recent studies of related quantum-dissipative
systems, such as the spin-boson model.~\cite{stauber+guinea} Furthermore, our findings
underscore the fact that the nature of the phonon-induced localization in the presence 
of nonlocal particle-phonon interactions is different than that of the purely local 
interactions. This may have bearing not only on the solid-state systems exhibiting 
polaronic behavior, but possibly also on certain classes of cold-atom
systems -- with (in principle) tunable couplings -- where phonons can be introduced in
a controlled way.~\cite{ColdPolaron} The need to investigate the interplay between
entanglement and the phonon-induced localization in other relevant models~\cite{Zhang+:08pre}
is clearly compelling.\vfill

\acknowledgments
M.V. acknowledges support by the Swiss National Science Foundation (SNSF).

\bibliography{polaronref}

\begin{thebibliography}{61}
\expandafter\ifx\csname natexlab\endcsname\relax\def\natexlab#1{#1}\fi
\expandafter\ifx\csname bibnamefont\endcsname\relax
  \def\bibnamefont#1{#1}\fi
\expandafter\ifx\csname bibfnamefont\endcsname\relax
  \def\bibfnamefont#1{#1}\fi
\expandafter\ifx\csname citenamefont\endcsname\relax
  \def\citenamefont#1{#1}\fi
\expandafter\ifx\csname url\endcsname\relax
  \def\url#1{\texttt{#1}}\fi
\expandafter\ifx\csname urlprefix\endcsname\relax\def\urlprefix{URL }\fi
\providecommand{\bibinfo}[2]{#2}
\providecommand{\eprint}[2][]{\url{#2}}

\bibitem[{Ple()}]{PlenioReview:07}
\bibinfo{note}{For an introduction, see M. B. Plenio and S. Virmani, Quantum
  Inf. Comput. $\mathbf{7}$, 1 (2007).}

\bibitem[{\citenamefont{Bengtsson and \.{Z}yczkowski}(2006)}]{EntangleBook}
\bibinfo{author}{\bibfnamefont{I.}~\bibnamefont{Bengtsson}} \bibnamefont{and}
  \bibinfo{author}{\bibfnamefont{K.}~\bibnamefont{\.{Z}yczkowski}},
  \emph{\bibinfo{title}{Geometry of {Q}uantum {S}tates: {A}n introduction to
  quantum entanglement}} (\bibinfo{publisher}{Cambridge University Press},
  \bibinfo{address}{Cambridge}, \bibinfo{year}{2006}).

\bibitem[{\citenamefont{Bennett}(1995)}]{Bennett:95}
\bibinfo{author}{\bibfnamefont{C.~H.} \bibnamefont{Bennett}},
  \bibinfo{journal}{Physics Today} \textbf{\bibinfo{volume}{48}},
  \bibinfo{pages}{24} (\bibinfo{year}{1995}).

\bibitem[{Ost()}]{Osterloh+Nielsen:02}
\bibinfo{note}{A. Osterloh, L. Amico, G. Falci, and R. Fazio, Nature (London)
  ${\mathbf{416}}$, $608$ ($2002$); T. J. Osborne and M. A. Nielsen, Phys. Rev.
  Lett. ${\mathbf{66}}$, $032110$ ($2002$).}

\bibitem[{\citenamefont{Galindo and Martin-Delgado}(2002)}]{Galindo:02}
\bibinfo{author}{\bibfnamefont{A.}~\bibnamefont{Galindo}} \bibnamefont{and}
  \bibinfo{author}{\bibfnamefont{M.~A.} \bibnamefont{Martin-Delgado}},
  \bibinfo{journal}{Rev. Mod. Phys.} \textbf{\bibinfo{volume}{74}}
  (\bibinfo{year}{2002}).

\bibitem[{\citenamefont{Wu et~al.}(2006)\citenamefont{Wu, Sarandy, Lidar, and
  Sham}}]{LidarSham:06}
\bibinfo{author}{\bibfnamefont{L.-A.} \bibnamefont{Wu}},
  \bibinfo{author}{\bibfnamefont{M.~S.} \bibnamefont{Sarandy}},
  \bibinfo{author}{\bibfnamefont{D.~A.} \bibnamefont{Lidar}}, \bibnamefont{and}
  \bibinfo{author}{\bibfnamefont{L.~J.} \bibnamefont{Sham}},
  \bibinfo{journal}{Phys. Rev. A} \textbf{\bibinfo{volume}{74}},
  \bibinfo{pages}{052335} (\bibinfo{year}{2006}).

\bibitem[{ved()}]{vedralovi}
\bibinfo{note}{V. Vedral, Nature (London) $\mathbf{425}$, 28 (2003); New J.
  Phys $\mathbf{6}$, 102 (2004); G. De Chiara, \v{C}. Brukner, R. Fazio, G. M.
  Palma, and V. Vedral, New J. Phys $\mathbf{8}$, 95 (2006).}

\bibitem[{Fac()}]{Factorization}
\bibinfo{note}{Entanglement measures may signify the occurrence of factorized
  quantum states which is not observed in conventional thermodynamic properties
  of the system. See, for example, T. Roscilde, P. Verrucchi, A. Fubini, S.
  Haas, and V. Tognetti, Phys. Rev. Lett. ${\mathbf{93}}$, 167203 ($2004$);
  ${\mathbf{94}}$, 147208 ($2005$); F. Baroni, A. Fubini, V. Tognetti, and P.
  Verrucchi, J. Phys. A: Math. Theor. ${\mathbf{40}}$, 9845 ($2007$); S. M.
  Giampaolo, G. Adesso, and F. Illuminati, Phys. Rev. Lett. ${\mathbf{100}}$,
  197201 ($2008$).}

\bibitem[{\citenamefont{Choi et~al.}(2000)\citenamefont{Choi, Bruder, and
  Loss}}]{Choi:00}
\bibinfo{author}{\bibfnamefont{M.-S.} \bibnamefont{Choi}},
  \bibinfo{author}{\bibfnamefont{C.}~\bibnamefont{Bruder}}, \bibnamefont{and}
  \bibinfo{author}{\bibfnamefont{D.}~\bibnamefont{Loss}},
  \bibinfo{journal}{Phys. Rev. B} \textbf{\bibinfo{volume}{62}},
  \bibinfo{pages}{13569} (\bibinfo{year}{2000}).

\bibitem[{mes()}]{mesoentangle}
\bibinfo{note}{G. Burkard, D. Loss, and E. V. Sukhorukov, Phys. Rev. B
  ${\mathbf{61}}$, R$16303$ ($2000$); N. M. Chtchelkatchev, G. Blatter, G. B.
  Lesovik, and T. Martin, Phys. Rev. B ${\mathbf{66}}$, $161320$(R) ($2002$);
  P. Samuelsson, E. V. Sukhorukov, and M. B{\"{u}}ttiker, Phys. Rev. Lett.
  ${\mathbf{91}}$, $157002$ ($2003$); M. Kindermann, Phys. Rev. Lett.
  ${\mathbf{96}}$, $240403$ ($2006$); I. Klich and L. Levitov,
  arXiv:0804.1377v2 (unpublished).}

\bibitem[{\citenamefont{Beenakker}(2006)}]{BeenakkerLec:06}
\bibinfo{author}{\bibfnamefont{C.~W.~J.} \bibnamefont{Beenakker}}, in
  \emph{\bibinfo{booktitle}{Proc. Int. School of Physics "E. Fermi", Vol.
  162}}, edited by \bibinfo{editor}{\bibfnamefont{G.}~\bibnamefont{Casati}},
  \bibinfo{editor}{\bibfnamefont{D.~L.} \bibnamefont{Shepelyansky}},
  \bibinfo{editor}{\bibfnamefont{P.}~\bibnamefont{Zoller}}, \bibnamefont{and}
  \bibinfo{editor}{\bibfnamefont{G.}~\bibnamefont{Benenti}}
  (\bibinfo{publisher}{IOS Press}, \bibinfo{address}{Amsterdam},
  \bibinfo{year}{2006}), pp. \bibinfo{pages}{307--347}.

\bibitem[{\citenamefont{Verstraete et~al.}(2004)\citenamefont{Verstraete,
  Martin-Delgado, and Cirac}}]{Verstraete:04}
\bibinfo{author}{\bibfnamefont{F.}~\bibnamefont{Verstraete}},
  \bibinfo{author}{\bibfnamefont{M.~A.} \bibnamefont{Martin-Delgado}},
  \bibnamefont{and} \bibinfo{author}{\bibfnamefont{J.~I.} \bibnamefont{Cirac}},
  \bibinfo{journal}{Phys. Rev. Lett.} \textbf{\bibinfo{volume}{92}},
  \bibinfo{pages}{087201} (\bibinfo{year}{2004}).

\bibitem[{\citenamefont{Garcia-Ripoll et~al.}(2004)\citenamefont{Garcia-Ripoll,
  Martin-Delgado, and Cirac}}]{Juancho:04}
\bibinfo{author}{\bibfnamefont{J.~J.} \bibnamefont{Garcia-Ripoll}},
  \bibinfo{author}{\bibfnamefont{M.~A.} \bibnamefont{Martin-Delgado}},
  \bibnamefont{and} \bibinfo{author}{\bibfnamefont{J.~I.} \bibnamefont{Cirac}},
  \bibinfo{journal}{Phys. Rev. Lett.} \textbf{\bibinfo{volume}{93}},
  \bibinfo{pages}{250405} (\bibinfo{year}{2004}).

\bibitem[{\citenamefont{Fan et~al.}(2004)\citenamefont{Fan, Korepin, and
  Roychowdhury}}]{Fan:04}
\bibinfo{author}{\bibfnamefont{H.}~\bibnamefont{Fan}},
  \bibinfo{author}{\bibfnamefont{V.}~\bibnamefont{Korepin}}, \bibnamefont{and}
  \bibinfo{author}{\bibfnamefont{V.}~\bibnamefont{Roychowdhury}},
  \bibinfo{journal}{Phys. Rev. Lett.} \textbf{\bibinfo{volume}{93}},
  \bibinfo{pages}{227203} (\bibinfo{year}{2004}).

\bibitem[{\citenamefont{Brand{\~{a}}o}(2005)}]{Brandao:05}
\bibinfo{author}{\bibfnamefont{F.~G. S.~L.} \bibnamefont{Brand{\~{a}}o}},
  \bibinfo{journal}{New {J}. {P}hys.} \textbf{\bibinfo{volume}{7}},
  \bibinfo{pages}{254} (\bibinfo{year}{2005}).

\bibitem[{\citenamefont{Larsson and Johannesson}(2005)}]{Larsson:05}
\bibinfo{author}{\bibfnamefont{D.}~\bibnamefont{Larsson}} \bibnamefont{and}
  \bibinfo{author}{\bibfnamefont{H.}~\bibnamefont{Johannesson}},
  \bibinfo{journal}{Phys. Rev. Lett.} \textbf{\bibinfo{volume}{95}},
  \bibinfo{pages}{196406} (\bibinfo{year}{2005}).

\bibitem[{\citenamefont{Vollbrecht and Cirac}(2007)}]{Vollbrecht:07}
\bibinfo{author}{\bibfnamefont{K.~G.~H.} \bibnamefont{Vollbrecht}}
  \bibnamefont{and} \bibinfo{author}{\bibfnamefont{J.~I.} \bibnamefont{Cirac}},
  \bibinfo{journal}{Phys. Rev. Lett.} \textbf{\bibinfo{volume}{98}},
  \bibinfo{pages}{190502} (\bibinfo{year}{2007}).

\bibitem[{\citenamefont{Ba{\~{n}}uls et~al.}(2007)\citenamefont{Ba{\~{n}}uls,
  Cirac, and Wolf}}]{Banuls:08}
\bibinfo{author}{\bibfnamefont{M.-C.} \bibnamefont{Ba{\~{n}}uls}},
  \bibinfo{author}{\bibfnamefont{J.~I.} \bibnamefont{Cirac}}, \bibnamefont{and}
  \bibinfo{author}{\bibfnamefont{M.~M.} \bibnamefont{Wolf}},
  \bibinfo{journal}{Phys. Rev. A} \textbf{\bibinfo{volume}{76}},
  \bibinfo{pages}{022311} (\bibinfo{year}{2007}).

\bibitem[{sta()}]{stauber+guinea}
\bibinfo{note}{T. Stauber and F. Guinea, Phys. Rev. A ${\mathbf{70}}$, $022313$
  ($2004$); ${\mathbf{73}}$, $042110$ ($2006$).}

\bibitem[{\citenamefont{Le\:Hur}(2008)}]{LeHur:08}
\bibinfo{author}{\bibfnamefont{K.}~\bibnamefont{Le\:Hur}},
  \bibinfo{journal}{Ann. Phys.} \textbf{\bibinfo{volume}{323}},
  \bibinfo{pages}{2208} (\bibinfo{year}{2008}).

\bibitem[{\citenamefont{Kopp et~al.}(2007)\citenamefont{Kopp, Jia, and
  Chakravarty}}]{Kopp:07}
\bibinfo{author}{\bibfnamefont{A.}~\bibnamefont{Kopp}},
  \bibinfo{author}{\bibfnamefont{X.}~\bibnamefont{Jia}}, \bibnamefont{and}
  \bibinfo{author}{\bibfnamefont{S.}~\bibnamefont{Chakravarty}},
  \bibinfo{journal}{Ann. Phys.} \textbf{\bibinfo{volume}{322}},
  \bibinfo{pages}{1466} (\bibinfo{year}{2007}).

\bibitem[{Ami()}]{Amico:08}
\bibinfo{note}{For an extensive review, see L. Amico, R. Fazio, A. Osterloh,
  and V. Vedral, Rev. Mod. Phys. $\bf{80}$, 517-576 (2008).}

\bibitem[{lan()}]{landaupekar}
\bibinfo{note}{L. D. Landau, Z. Phys. ${\mathbf {3}}$, $664$ ($1933$); L. D.
  Landau and S. I. Pekar, Zh. Eksp. Teor. Fiz. ${\mathbf {18}}$, $419$
  ($1948$).}

\bibitem[{\citenamefont{Firsov}(1975)}]{firsovbook}
\bibinfo{author}{\bibfnamefont{Y.~A.} \bibnamefont{Firsov}},
  \emph{\bibinfo{title}{Polarons}} (\bibinfo{publisher}{Mir},
  \bibinfo{address}{Moscow}, \bibinfo{year}{1975}).

\bibitem[{\citenamefont{Alexandrov and Mott}(1995)}]{AlexandrovBook}
\bibinfo{author}{\bibfnamefont{A.~S.} \bibnamefont{Alexandrov}}
  \bibnamefont{and} \bibinfo{author}{\bibfnamefont{N.}~\bibnamefont{Mott}},
  \emph{\bibinfo{title}{Polarons and Bipolarons}} (\bibinfo{publisher}{World
  Scientific}, \bibinfo{address}{Singapore}, \bibinfo{year}{1995}).

\bibitem[{Pol()}]{PolaronReview}
\bibinfo{note}{J. Ranninger, in {\em Proc. Int. School of Physics ``E. Fermi'',
  Course CLXI}, edited by G. Iadonisi, J. Ranninger, and G. De\:Filippis (IOS
  Press, Amsterdam, 2006), pp. 1--25; O. S. Bari{\v{s}}i\'c and S.
  Bari{\v{s}}i\'c, Eur. Phys. J. B ${\mathbf{64}}$, $1$ ($2008$).}

\bibitem[{pol({\natexlab{a}})}]{polaroncrucial}
\bibinfo{note}{T. Holstein, Ann. Phys. ${\mathbf {8}}$, 343 (1959); I. G. Lang
  and Yu. A. Firsov, Zh. Eksp. Teor. Fiz. ${\mathbf {43}}$, 1843 (1962) [ Sov.
  Phys. JETP ${\mathbf {16}}$, 1301 (1963)]; J. Ranninger and U. Thibblin,
  Phys. Rev. B ${\mathbf{45}}$, 7730 (1992); G. Wellein and H. Fehske, Phys.
  Rev. B ${\mathbf{56}}$, 4513 (1997).}

\bibitem[{Col()}]{ColdPolaron}
\bibinfo{note}{I. E. Mazets, G. Kurizki, N. Katz, and N. Davidson, Phys. Rev.
  Lett. ${\mathbf{94}}$, $190403$ ($2005$); E. Pazy and A. Vardi, Phys. Rev. A
  ${\mathbf{72}}$, $033609$ ($2005$); K. G{\"{u}}nter, T. St{\"{o}}ferle, H.
  Moritz, M. K{\"{o}}hl, and T. Esslinger, Phys. Rev. Lett. ${\mathbf{96}}$,
  $180402$ ($2006$); F. M. Cucchietti and E. Timmermans, Phys. Rev. Lett.
  ${\mathbf{96}}$, $210401$ ($2006$); L. Mathey and D.-W. Wang, Phys. Rev. A
  ${\mathbf{75}}$, $013612$ ($2007$); M. Bruderer, A. Klein, S. R. Clark, and
  D. Jaksch, Phys. Rev. A ${\mathbf{76}}$, $011605$(R) ($2007$); L. Pollet, C.
  Kollath, U. Schollw{\"{o}}ck, and M. Troyer, Phys. Rev. A ${\mathbf{77}}$,
  $023608$ ($2008$).}

\bibitem[{pol({\natexlab{b}})}]{polaronhtc}
\bibinfo{note}{S. Ishihara and N. Nagaosa, Phys. Rev. B ${\mathbf {69}}$,
  $144520$ ($2004$); O. R{\"{o}}sch and O. Gunnarsson, Phys. Rev. Lett.
  ${\mathbf {92}}$, $146403$ ($2004$); C. Slezak, A. Macridin, G. A. Sawatzky,
  M. Jarrell, and T. A. Maier, Phys. Rev. B ${\mathbf {73}}$, $205122$
  ($2006$).}

\bibitem[{\citenamefont{Perroni et~al.}(2004)\citenamefont{Perroni, Piegari,
  Capone, and Cataudella}}]{Perroni:04}
\bibinfo{author}{\bibfnamefont{C.~A.} \bibnamefont{Perroni}},
  \bibinfo{author}{\bibfnamefont{E.}~\bibnamefont{Piegari}},
  \bibinfo{author}{\bibfnamefont{M.}~\bibnamefont{Capone}}, \bibnamefont{and}
  \bibinfo{author}{\bibfnamefont{V.}~\bibnamefont{Cataudella}},
  \bibinfo{journal}{Phys. Rev. B} \textbf{\bibinfo{volume}{69}},
  \bibinfo{pages}{174301} (\bibinfo{year}{2004}).

\bibitem[{\citenamefont{Perroni et~al.}(2005)\citenamefont{Perroni, Cataudella,
  De\:Filippis, and Ramaglia}}]{Perroni:05}
\bibinfo{author}{\bibfnamefont{C.~A.} \bibnamefont{Perroni}},
  \bibinfo{author}{\bibfnamefont{V.}~\bibnamefont{Cataudella}},
  \bibinfo{author}{\bibfnamefont{G.}~\bibnamefont{De\:Filippis}},
  \bibnamefont{and} \bibinfo{author}{\bibfnamefont{V.~M.}
  \bibnamefont{Ramaglia}}, \bibinfo{journal}{Phys. Rev. B}
  \textbf{\bibinfo{volume}{71}}, \bibinfo{pages}{054301}
  (\bibinfo{year}{2005}).

\bibitem[{\citenamefont{Zaanen and Littlewood}(1994)}]{Zaanen:94}
\bibinfo{author}{\bibfnamefont{J.}~\bibnamefont{Zaanen}} \bibnamefont{and}
  \bibinfo{author}{\bibfnamefont{P.~B.} \bibnamefont{Littlewood}},
  \bibinfo{journal}{Phys. Rev. B} \textbf{\bibinfo{volume}{50}},
  \bibinfo{pages}{7222} (\bibinfo{year}{1994}).

\bibitem[{\citenamefont{Zaanen}(1996)}]{zaanendictaat}
\bibinfo{author}{\bibfnamefont{J.}~\bibnamefont{Zaanen}},
  \emph{\bibinfo{title}{The {C}lassical {C}ondensates: From {C}rystals to
  {F}ermi-liquids}} (\bibinfo{publisher}{Lorentz Institute for Theoretical
  Physics}, \bibinfo{address}{Leiden}, \bibinfo{year}{1996}).

\bibitem[{\citenamefont{Yonemitsu and Maeshima}(2007)}]{yonemitsu:07}
\bibinfo{author}{\bibfnamefont{K.}~\bibnamefont{Yonemitsu}} \bibnamefont{and}
  \bibinfo{author}{\bibfnamefont{N.}~\bibnamefont{Maeshima}},
  \bibinfo{journal}{Phys. Rev. B} \textbf{\bibinfo{volume}{76}},
  \bibinfo{pages}{075105} (\bibinfo{year}{2007}).

\bibitem[{SSH()}]{SSH}
\bibinfo{note}{W. P. Su, J. R. Schrieffer, and A. J. Heeger, Phys. Rev. Lett.
  ${\mathbf {42}}$, $1698$ ($1979$); A. J. Heeger, S. Kivelson, J. R.
  Schrieffer, and W. P. Su, Rev. Mod. Phys. ${\mathbf {60}}$, $781$ ($1988$).}

\bibitem[{\citenamefont{Zoli}(2007)}]{ZoliIn}
\bibinfo{author}{\bibfnamefont{M.}~\bibnamefont{Zoli}}, in
  \emph{\bibinfo{booktitle}{Polarons in Advanced Materials}}, edited by
  \bibinfo{editor}{\bibfnamefont{A.~S.} \bibnamefont{Alexandrov}}
  (\bibinfo{publisher}{Canopus Books}, \bibinfo{address}{Bristol},
  \bibinfo{year}{2007}).

\bibitem[{zol()}]{zolicouple}
\bibinfo{note}{M. Zoli, Phys. Rev. B ${\mathbf {67}}$, $195102$ ($2003$);
  ${\mathbf {70}}$, $184301$ ($2004$); ${\mathbf {71}}$, $205111$ ($2005$).}

\bibitem[{\citenamefont{Cross and Fisher}(1979)}]{Cross:79}
\bibinfo{author}{\bibfnamefont{M.~C.} \bibnamefont{Cross}} \bibnamefont{and}
  \bibinfo{author}{\bibfnamefont{D.~S.} \bibnamefont{Fisher}},
  \bibinfo{journal}{Phys. Rev. B} \textbf{\bibinfo{volume}{19}},
  \bibinfo{pages}{402} (\bibinfo{year}{1979}).

\bibitem[{\citenamefont{Barford and Bursill}(2005)}]{Barford:05}
\bibinfo{author}{\bibfnamefont{W.}~\bibnamefont{Barford}} \bibnamefont{and}
  \bibinfo{author}{\bibfnamefont{R.~J.} \bibnamefont{Bursill}},
  \bibinfo{journal}{Phys. Rev. Lett.} \textbf{\bibinfo{volume}{95}},
  \bibinfo{pages}{137207} (\bibinfo{year}{2005}).

\bibitem[{\citenamefont{Hannewald et~al.}(2004)\citenamefont{Hannewald,
  Stojanovi\'{c}, Schellekens, Bobbert, Kresse, and Hafner}}]{hannewald:04}
\bibinfo{author}{\bibfnamefont{K.}~\bibnamefont{Hannewald}},
  \bibinfo{author}{\bibfnamefont{V.~M.} \bibnamefont{Stojanovi\'{c}}},
  \bibinfo{author}{\bibfnamefont{J.~M.~T.} \bibnamefont{Schellekens}},
  \bibinfo{author}{\bibfnamefont{P.~A.} \bibnamefont{Bobbert}},
  \bibinfo{author}{\bibfnamefont{G.}~\bibnamefont{Kresse}}, \bibnamefont{and}
  \bibinfo{author}{\bibfnamefont{J.}~\bibnamefont{Hafner}},
  \bibinfo{journal}{Phys. Rev. B} \textbf{\bibinfo{volume}{69}},
  \bibinfo{pages}{075211} (\bibinfo{year}{2004}).

\bibitem[{\citenamefont{Stojanovi\'{c}
  et~al.}(2004)\citenamefont{Stojanovi\'{c}, Bobbert, and
  Michels}}]{stojanovic:04}
\bibinfo{author}{\bibfnamefont{V.~M.} \bibnamefont{Stojanovi\'{c}}},
  \bibinfo{author}{\bibfnamefont{P.~A.} \bibnamefont{Bobbert}},
  \bibnamefont{and} \bibinfo{author}{\bibfnamefont{M.~A.~J.}
  \bibnamefont{Michels}}, \bibinfo{journal}{Phys. Rev. B}
  \textbf{\bibinfo{volume}{69}}, \bibinfo{pages}{144302}
  (\bibinfo{year}{2004}).

\bibitem[{\citenamefont{FoaTorres and Roche}(2006)}]{Torres:06}
\bibinfo{author}{\bibfnamefont{L.~E.~F.} \bibnamefont{FoaTorres}}
  \bibnamefont{and} \bibinfo{author}{\bibfnamefont{S.}~\bibnamefont{Roche}},
  \bibinfo{journal}{Phys. Rev. Lett.} \textbf{\bibinfo{volume}{97}},
  \bibinfo{pages}{076804} (\bibinfo{year}{2006}).

\bibitem[{\citenamefont{Schmidt et~al.}(2007)\citenamefont{Schmidt, Hettler,
  and Sch{\"{o}}n}}]{dnaschoen:07}
\bibinfo{author}{\bibfnamefont{B.~B.} \bibnamefont{Schmidt}},
  \bibinfo{author}{\bibfnamefont{M.~H.} \bibnamefont{Hettler}},
  \bibnamefont{and}
  \bibinfo{author}{\bibfnamefont{G.}~\bibnamefont{Sch{\"{o}}n}},
  \bibinfo{journal}{Phys. Rev. B} \textbf{\bibinfo{volume}{75}},
  \bibinfo{pages}{115125} (\bibinfo{year}{2007}).

\bibitem[{\citenamefont{Alvermann et~al.}(2007)\citenamefont{Alvermann,
  Edwards, and Fehske}}]{alvermann:07}
\bibinfo{author}{\bibfnamefont{A.}~\bibnamefont{Alvermann}},
  \bibinfo{author}{\bibfnamefont{D.~M.} \bibnamefont{Edwards}},
  \bibnamefont{and} \bibinfo{author}{\bibfnamefont{H.}~\bibnamefont{Fehske}},
  \bibinfo{journal}{Phys. Rev. Lett.} \textbf{\bibinfo{volume}{98}},
  \bibinfo{pages}{056602} (\bibinfo{year}{2007}).

\bibitem[{Ger()}]{Gerlach:87}
\bibinfo{note}{B. Gerlach and H. Lowen, Phys. Rev. B ${\mathbf{35}}$, $4291$
  ($1987$); ${\mathbf{35}}$, $4297$ ($1987$).}

\bibitem[{\citenamefont{Zhao et~al.}(2004)\citenamefont{Zhao, Zanardi, and
  Chen}}]{zhaozanardi}
\bibinfo{author}{\bibfnamefont{Y.}~\bibnamefont{Zhao}},
  \bibinfo{author}{\bibfnamefont{P.}~\bibnamefont{Zanardi}}, \bibnamefont{and}
  \bibinfo{author}{\bibfnamefont{G.}~\bibnamefont{Chen}},
  \bibinfo{journal}{Phys. Rev. B} \textbf{\bibinfo{volume}{70}},
  \bibinfo{pages}{195113} (\bibinfo{year}{2004}).

\bibitem[{\citenamefont{Porras et~al.}(2008)\citenamefont{Porras, Marquardt,
  von Delft, and Cirac}}]{Porras:08}
\bibinfo{author}{\bibfnamefont{D.}~\bibnamefont{Porras}},
  \bibinfo{author}{\bibfnamefont{F.}~\bibnamefont{Marquardt}},
  \bibinfo{author}{\bibfnamefont{J.}~\bibnamefont{von Delft}},
  \bibnamefont{and} \bibinfo{author}{\bibfnamefont{J.~I.} \bibnamefont{Cirac}},
  \bibinfo{journal}{Phys. Rev. A} \textbf{\bibinfo{volume}{78}},
  \bibinfo{pages}{010101(R)} (\bibinfo{year}{2008}).

\bibitem[{\citenamefont{Wei{\ss}e et~al.}(2000)\citenamefont{Wei{\ss}e, Fehske,
  Wellein, and Bishop}}]{Weisse:00}
\bibinfo{author}{\bibfnamefont{A.}~\bibnamefont{Wei{\ss}e}},
  \bibinfo{author}{\bibfnamefont{H.}~\bibnamefont{Fehske}},
  \bibinfo{author}{\bibfnamefont{G.}~\bibnamefont{Wellein}}, \bibnamefont{and}
  \bibinfo{author}{\bibfnamefont{A.~R.} \bibnamefont{Bishop}},
  \bibinfo{journal}{Phys. Rev. B} \textbf{\bibinfo{volume}{62}},
  \bibinfo{pages}{R747} (\bibinfo{year}{2000}).

\bibitem[{\citenamefont{Ku et~al.}(2002)\citenamefont{Ku, Trugman, and
  Bon\v{c}a}}]{Ku+Bonca:02}
\bibinfo{author}{\bibfnamefont{L.-C.} \bibnamefont{Ku}},
  \bibinfo{author}{\bibfnamefont{S.~A.} \bibnamefont{Trugman}},
  \bibnamefont{and}
  \bibinfo{author}{\bibfnamefont{J.}~\bibnamefont{Bon\v{c}a}},
  \bibinfo{journal}{Phys. Rev. B} \textbf{\bibinfo{volume}{65}},
  \bibinfo{pages}{174306} (\bibinfo{year}{2002}).

\bibitem[{\citenamefont{Toyozawa}(1961)}]{Toyozawa:61}
\bibinfo{author}{\bibfnamefont{Y.}~\bibnamefont{Toyozawa}},
  \bibinfo{journal}{Prog. Theor. Phys.} \textbf{\bibinfo{volume}{26}},
  \bibinfo{pages}{29} (\bibinfo{year}{1961}).

\bibitem[{Rom()}]{Romero+Lindenberg}
\bibinfo{note}{A. H. Romero, D. W. Brown, and K. Lindenberg, J. Chem. Phys.
  ${\mathbf{109}}$, $6540$ ($1998$); Phys. Rev. B ${\mathbf{59}}$, $13728$
  ($1999$).}

\bibitem[{\citenamefont{Bari\v{s}i\'{c}}(2002)}]{Barisic:02}
\bibinfo{author}{\bibfnamefont{O.~S.} \bibnamefont{Bari\v{s}i\'{c}}},
  \bibinfo{journal}{Phys. Rev. B} \textbf{\bibinfo{volume}{65}},
  \bibinfo{pages}{144301} (\bibinfo{year}{2002}).

\bibitem[{\citenamefont{Scully and Zubairy}(1997)}]{Scully+Zubairy}
\bibinfo{author}{\bibfnamefont{M.~O.} \bibnamefont{Scully}} \bibnamefont{and}
  \bibinfo{author}{\bibfnamefont{M.~S.} \bibnamefont{Zubairy}},
  \emph{\bibinfo{title}{Quantum {O}ptics}} (\bibinfo{publisher}{Cambridge
  University Press}, \bibinfo{address}{Cambridge}, \bibinfo{year}{1997}).

\bibitem[{\citenamefont{T{\"{o}}rn and {\v{Z}}ilinskas}(1989)}]{OptimBook}
\bibinfo{author}{\bibfnamefont{A.}~\bibnamefont{T{\"{o}}rn}} \bibnamefont{and}
  \bibinfo{author}{\bibfnamefont{A.}~\bibnamefont{{\v{Z}}ilinskas}},
  \emph{\bibinfo{title}{Global {O}ptimization}} (\bibinfo{publisher}{Springer},
  \bibinfo{address}{New York}, \bibinfo{year}{1989}).

\bibitem[{\citenamefont{Cataudella et~al.}(1999)\citenamefont{Cataudella,
  De\:Filippis, and Iadonisi}}]{Cataudella:99}
\bibinfo{author}{\bibfnamefont{V.}~\bibnamefont{Cataudella}},
  \bibinfo{author}{\bibfnamefont{G.}~\bibnamefont{De\:Filippis}},
  \bibnamefont{and} \bibinfo{author}{\bibfnamefont{G.}~\bibnamefont{Iadonisi}},
  \bibinfo{journal}{Phys. Rev. B} \textbf{\bibinfo{volume}{60}},
  \bibinfo{pages}{15163} (\bibinfo{year}{1999}).

\bibitem[{\citenamefont{Jeckelmann and White}(1998)}]{Jeckelmann:98}
\bibinfo{author}{\bibfnamefont{E.}~\bibnamefont{Jeckelmann}} \bibnamefont{and}
  \bibinfo{author}{\bibfnamefont{S.~R.} \bibnamefont{White}},
  \bibinfo{journal}{Phys. Rev. B} \textbf{\bibinfo{volume}{57}},
  \bibinfo{pages}{6376} (\bibinfo{year}{1998}).

\bibitem[{Pol()}]{PolaronQMC}
\bibinfo{note}{H. De Raedt and A. Lagendijk, Phys. Rev. B ${\mathbf{27}}$,
  $6097$ ($1983$); ${\mathbf{30}}$, $1671$ ($1984$); P. E. Kornilovitch and E.
  R. Pike, Phys. Rev. B ${\mathbf{55}}$, R$8634$ ($1997$).}

\bibitem[{\citenamefont{Sachdev}(1999)}]{sachdevbook}
\bibinfo{author}{\bibfnamefont{S.}~\bibnamefont{Sachdev}},
  \emph{\bibinfo{title}{Quantum {P}hase {T}ransitions}}
  (\bibinfo{publisher}{Cambridge University Press}, \bibinfo{address}{New
  York}, \bibinfo{year}{1999}).

\bibitem[{\citenamefont{Shore and Sander}(1973)}]{Shore:73}
\bibinfo{author}{\bibfnamefont{H.~B.} \bibnamefont{Shore}} \bibnamefont{and}
  \bibinfo{author}{\bibfnamefont{L.~M.} \bibnamefont{Sander}},
  \bibinfo{journal}{Phys. Rev. B} \textbf{\bibinfo{volume}{7}},
  \bibinfo{pages}{4537} (\bibinfo{year}{1973}).

\bibitem[{\citenamefont{Chen and Wong}(2008)}]{Chen+Wong:08pre}
\bibinfo{author}{\bibfnamefont{Z.-D.} \bibnamefont{Chen}} \bibnamefont{and}
  \bibinfo{author}{\bibfnamefont{H.}~\bibnamefont{Wong}},
  \bibinfo{journal}{Phys. Rev. B} \textbf{\bibinfo{volume}{78}},
  \bibinfo{pages}{064308} (\bibinfo{year}{2008}).

\bibitem[{\citenamefont{Brand et~al.}(2008)\citenamefont{Brand, Flach, Fleurov,
  Schulman, and Tolkunov}}]{Brand:08}
\bibinfo{author}{\bibfnamefont{J.}~\bibnamefont{Brand}},
  \bibinfo{author}{\bibfnamefont{S.}~\bibnamefont{Flach}},
  \bibinfo{author}{\bibfnamefont{V.}~\bibnamefont{Fleurov}},
  \bibinfo{author}{\bibfnamefont{L.~S.} \bibnamefont{Schulman}},
  \bibnamefont{and} \bibinfo{author}{\bibfnamefont{D.}~\bibnamefont{Tolkunov}},
  \bibinfo{journal}{Europhys. Lett.} \textbf{\bibinfo{volume}{83}},
  \bibinfo{pages}{40002} (\bibinfo{year}{2008}).

\end{thebibliography}
\bibliographystyle{apsrev}

\end{document}